# Does Explanation Matter? An Exploratory Study on the Effects of Covid–19 Misinformation Warning Flags on Social Media

Dipto Barman
School of Computer Science and Statistics
Trinity College Dublin
Dublin, Ireland
barmand@tcd.ie

Owen Colan
School of Computer Science and Statistics
Trinity College Dublin
Dublin, Ireland
owen.conlan@tcd.ie

*Abstract*— Digital platforms have employed flagging techniques to tackle misinformation as they offer a promising means of informing users about harmful content without resorting to censorship. However, their effectiveness depends on the user's understanding of the flags. Fact-checkers have been crucial in tackling misinformation online, but interestingly, fact-checked explanations have rarely been incorporated directly into the warning flags. They have usually been linked and directed towards their websites. Therefore, this study investigates user responses to misinformation flags in a hypothetical social media setting. It focuses on whether warnings influence users' perceived accuracy judgement and sharing intent of the false headlines. We also investigate whether adding specific explanations from fact-checking websites enhances trust in these flags. We conducted an experiment with 348 American participants, exposing them to a randomised order of true and false news headlines related to COVID-19, with and without warning flags and explanation text. Our findings suggest that warning flags, whether alone or accompanied by explanatory text, effectively reduce the perceived accuracy of fake news and the intent to share such headlines. Interestingly, our study also suggests that incorporating explanatory text in misinformation warning systems could significantly enhance their trustworthiness, emphasising the importance of transparency and user comprehension in combating fake news on social media.

## I. Introduction

The advent of social media has allowed various opinions and ideas to coexist. However, due to the increase in the affordance of digital platforms, they have become a source of false and misleading information online. Users have become prone to consume misinformation, disinformation, propaganda, and conspiracy theories. These platforms often have both verified and unverified claims appearing side-by-side. Misinformation is defined as "False or misleading information" [1]. Termed an "Infodemic" [2] by the World Health Organization (WHO), it has left many individuals confused about what exactly the truth is. In recent times, there have been numerous instances of misinformation related to the COVID-19 vaccines, including false claims about the safety and effectiveness of the vaccines, conspiracy theories about their development, and efforts to spread misinformation about the vaccine distribution process [3].

To effectively combat the spread of misinformation on social media while maintaining a balance between necessary moderation and excessive censorship, major digital platforms like Twitter and Facebook have introduced various mechanisms to label or flag content identified as potentially false or misleading [4]. Facebook and Twitter have been actively debunking misinformation online regarding COVID-19, vaccination and other false health-related information [4], [5] with fact checkers' help. This measure has been implemented to instantly alert users to the credibility of the information they come across, helping them make informed decisions about what they believe and share. Previous studies have shown flagging can reduce user susceptibility to misinformation [5]–[7].

However, these research studies also highlight the enormous design space for such warning flags, wherein the effectiveness could depend on various factors such as symbol choice [8], bot flags [6], crowd-sourced flagging [5] and content and source of the label [9]. A crucial aspect of these flagging systems is how the platforms attach warning flags or labels to specific posts. These flags typically involve visually distinguishable marks or notices attached to posts or links which have been identified, often through automated fact-checking mechanisms or reports from users, as containing information that is unverified, disputed, or outright false [8].

While flags alert users to potential misinformation, they do not provide context or counterarguments. This is where 'inoculation theory' [10], a concept from the field of psychology becomes relevant. Inoculation theory aims to expose the user to a weakened dose (i.e., explaining why certain information is false) of a misinformation argument so that the individual can immunise and confer resistance against the misinformation. Fact-checkers play a crucial role in this process, scrutinising the claims made in headlines for inconsistencies. However, these claims are usually presented on a separate fact-checker website rather than directly associated with the flagged content. This necessitates an additional cognitive step for users, requiring them to click on a link provided by the fact-checkers – an effort many users often skip [11].

Therefore, to reduce the cognitive load on users, our study focuses on a simulated environment of social media posts, some of which are tagged with misinformation flags. For our experiment, we supplemented some of the flags with explanatory text taken directly from fact-checkers' websites that refute the claims made in the headlines. The rationale behind our experiment lies in the intersection of the effectiveness of warning flags and the power of explanatory text in mitigating the effect of misinformation on users. By integrating explanatory text into the flagging mechanism, we aim to provide this missing context. The explanatory text acts as an immediate refutation to the misinformation. This enables users to understand not only that a piece of information has been flagged as potentially misleading but also why this is the case. We hypothesise that this would enhance the transparency of the flagging system and, in doing so, will foster greater trust and understanding among users. Drawing on research from the field of psychology, we anticipate that providing a direct refutation to the misinformation [10] within the flag itself will facilitate users in rejecting the misinformation's claims. This approach is not merely about identifying misinformation but also empowering users with the tools to dissect and challenge it.

In our study, we ask participants to rate the perceived accuracy and the sharing intent of each headline they encounter. We also ask participants to rate their perceived trust in the two flagging conditions (with just the misinformation flag and with context about the misinformation flag). This 'accuracy rating' and 'sharing intent' measure the perceived truthfulness and sharing intent of the news headline.

Considering these observations and gaps in current research, we investigate these research questions:

(i) whether these warnings in the form of flags reduce the overall accuracy of fake news items and sharing intent in a social media setting

(ii) whether adding explanations for why certain information is false, sourced directly from fact-checking websites, increases trust in the flagging system and

(iii) whether there is a correlation between various users' demographics (e.g., age, gender, education level, and political ideology) and their responses to different flagging systems.

By addressing these questions, we aim to provide valuable insights into designing more personalised and effective strategies that may be suitable for combating the spread of misinformation on digital platforms.

## II. Related Works

### A. Online Misinformation on Covid

Online misinformation can be modelled as a process that includes different actors and successive stages [12]. The model consists of bad actors (misinformation creators) who produce and push misleading content onto social media platforms that enable low-cost distribution and promotion of this content and the audience who consume and spread this information without any consequences.

Exposure to false information has been linked to negative impacts on society, such as promotions of anti-vaccination campaigns [13]. For example, recent research around COVID-19 misinformation [14] has indicated that people are less likely to follow public health guidelines [15] and have reduced intentions to get vaccinated and recommend vaccines to others [16]. This consumption of ambiguous information can further lead to life-threatening complications [17]. This has pressured researchers and social media companies such as Facebook and Twitter to develop methods to tackle online misinformation.

### B. Misinformation Flagging

Misinformation flagging has been a popular debunking technique employed by digital platforms to tackle misinformation without resorting to content moderation. Most debunking techniques are fact-based, but they can also appeal to logic and critical thinking, for example, by exposing a fallacious argumentation technique or the source of false information. Research has been done on using fact-checking and warning labels as refutational interventions in search engines and social media [6], [18]. Such flags can take different forms [8]. Some platforms opt to use blunt "false information" labels.

In contrast, others may provide a more nuanced "disputed" or "unverified" label, sometimes accompanied by a link to more reliable information or a fact-checking report. Several studies support the effectiveness of this approach, indicating that the application of warning flags to posts has reduced users' likelihood to believe and further disseminate the flagged content [5]–[7]. This is presumably because the presence of such warnings cues users to question the credibility of the information. However, these studies do not discuss whether individuals trust the flags or if it motivates them to seek out more reliable sources before accepting or sharing the content.

Currently, these flagging systems on digital platforms typically link to external fact-checking websites instead of incorporating the fact-checking information directly within the social media platform. This requires an additional cognitive step for users, who must navigate to another page to understand why the content was flagged, a step many users may not take [11].

Results from explainable recommendation studies suggest that adding explanations to recommendation systems online may increase the rate at which people accept that recommendation [19]. For example in [20], the author found that adding an explanation of how an AI system functions increases the warning's effectiveness to the user. However, they did not find an increase in self-reported trust in the warning label. It should also be noted that individuals tend to trust less content generated by AI-automated tools than flagging attributed by humans [21].

Thus, when designing misinformation warning systems, it is critical to consider not just whether the content is false or who is flagging it but also why it is deemed as such. We aim to bridge this gap by providing users with an explanation of the reason behind the flagging; we hypothesise that the credibility of the flagging system may be enhanced [22]. This, in turn, could increase the user's understanding and acceptance of the flags, making them more likely to question the credibility of flagged information and seek more reliable sources. However, despite the potential benefits, incorporating explanatory text into warning flags has yet to be extensively explored in existing research, presenting a significant gap that our study aims to address.

## III. METHODS

This study was conducted online using the Qualtrics online research platform. We recruited 384 American participants for our study using the Prolific platform, which employs quota matching to ensure an equal ratio of male and female participants. Before participant recruitment, this study was approved by the *omitted for submission* Ethics Committee. Out of the initial 384 participants, N = 10 participants completed the questionnaire too quickly and were subsequently excluded from the final data analysis. Among these participants, N = 15 failed the two attention check questions, N = 2 didn't complete the entire questionnaire, and N = 1 didn't consent to data collection at the end of the survey. Finally, after removing missing values in the records (N = 8), our final sample consisted of N = 348. Before viewing the stimuli, participants were asked about their gender (male, female, non-binary/third gender, Prefer not to say), age (18-25, 26-35, 36-45, 46-55, 56-65, 65+), education levels (less than high school, high school, undergraduate degree, graduate degree, post-graduate), social media usage per day (0-1, 2-3, 4-5, 5-6, 6+ hours) and political ideology (1 – extremely liberal to 5 – Extremely conservative). The final sample was 48.3% male and 48.9% female, with a mean age range of 26-35, a mean education level equivalent to an undergraduate degree, an average of 2-3 hours of social media use per day, and a political ideology skewing towards moderate liberalism. This is illustrated in Fig. 1.

This study used a within-subject design to mimic a social media feed. The headlines utilised in this experiment were selected from a larger set of headlines based on COVID-19 from an American perspective [23]. From a pool of 30 COVID-19-related headlines, 10 headlines were randomly selected and checked for topical relevance. To mitigate any potential bias from the source, we intentionally excluded the source website of the headline from the stimuli.

Participants were presented with three types of stimuli. They were:

i. No flag on the headline (control condition); these include two false headlines and three true headlines.

ii. A flagging condition in which participants were just shown "a fact-checker disputes the claim" (fake news headline with warning flag); these include two false headlines with flags.

iii. An explanation flagging condition where participants were shown "A fact-checker disputes the claim" and an explanation of why this claim is false along with the fact-checking website link. This piece of text was taken directly from the fact-checking website where the claim was being refuted (fake news headline with warning and explanation flags); these include three false headlines with warning and explanation flags.

Representations of the three stimulus types can be found in Fig. 2.

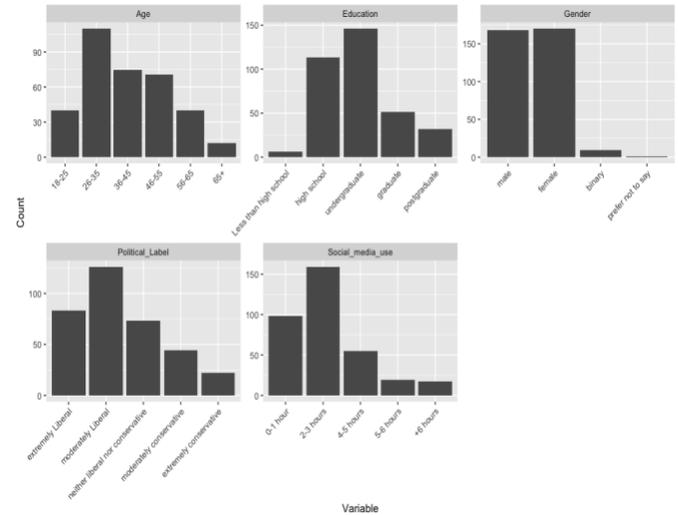

Fig. 1. Demographic variables for our sample.

The participants were shown these ten news headline stimuli in random order and asked to rate their accuracy ratings ("Given the presentation, how accurate do you think the headline is?") on a Likert scale from 1 (Not accurate at all) to 5 (very accurate). They were also asked to rate their intent about the headlines ("Given the presentation above, how likely are you to share this headline with your friends and family on social media?") on a Likert scale from 1 (Not likely at all) to 5 (very likely). In the flag conditions, we asked participants to rate the trustworthiness of the flags ("How trustworthy is the warning label and the associated text to you?") on a Likert scale from 1 (very untrustworthy) to 5 (very trustworthy). Two attention check questions were also inserted in a random order in the survey.

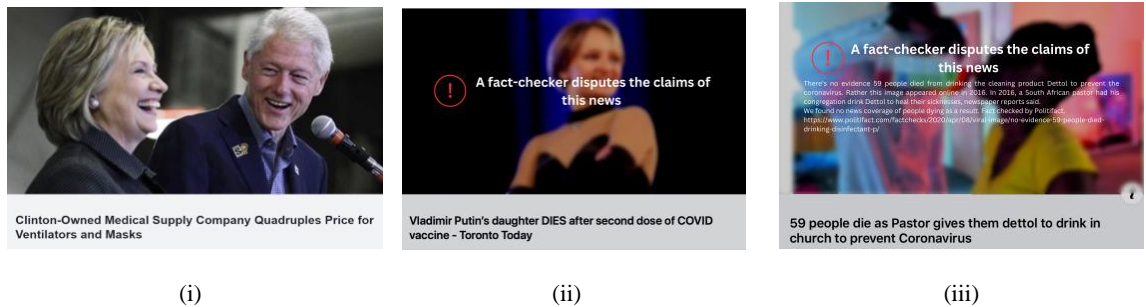

Fig. 2. The different stimuli's used in the experiment. (i) illustrates a fake headline (ii) illustrates a fake headline with a warning flag, and (iii) illustrates a fake headline with a warning and explanation flag.

IV. ANALYSIS AND RESULTS

*A. Analysis*

Our independent variables are age, education, gender, social media use, and political ideology. Our dependent variables are the perceived accuracy rating and sharing likelihoods of true and fake headlines, fake headlines with warning flags and fake headlines with warning and explanation flags. In the flagging conditions, we also measured the trustworthiness of the flags as a dependent variable. We used R statistical software to analyse our data. Using Kolmogorov-Smirnov and Shapiro-Wilk tests, we found that our data is not normally distributed. Consequently, for the purpose of this analysis, we chose to use non-parametric tests for further analysis. We employed the Friedman test to examine differences in accuracy and sharing likelihood ratings across the three types of stimuli and to compare the trustworthiness of the flagged conditions (i.e., fake news headline with warning flag vs fake news headline with warning and explanation flag) among the participants. To identify the correlation between the independent variables and dependent variables, we use Spearman's rank correlation.

*B. Results*

Regarding the Friedman test results, both perceived accuracy ratings (Friedman chi-squared = 441.77, p-value < 2.2e-16) and sharing likelihood rating (Friedman chi-squared = 203.24, p-value < 2.2e-16) have $p < 0.05$, indicating significant differences between at least two of the stimulus types. The result of the Friedman test for trustworthiness ratings indicates that there is a significant difference between the trustworthiness ratings of the two flagged conditions (Friedman chi-squared = 13.337, p-value = 0.0002602). This suggests that the presence or absence of the warning text has a significant impact on participants' trustworthiness ratings. Therefore, we use the Wilcoxon signed-rank tests with Bonferroni correction post hoc test for multiple comparisons, which will help identify which stimulus types have significant differences in ratings.

*1) Accuracy*

Using the Wilcoxon signed-rank tests, we find that overall participants rate true news headlines as more accurate compared to false news headlines. However, we also find that without the flags, participants rate fake news headlines as more accurate indicating that the flagging conditions do inherently reduce the accuracy intent of the news headline. However, in the flagging conditions, we find that fake news headlines with warning and explanation flags are rated as more accurate compared to fake news headlines with warning flags (p-value = 0.0491). This indicates that having explanation text with the warning flag does not significantly reduce the perceived accuracy of the fake news headline when compared to fake news headlines with only a warning flag. To show the effects of the flags, a box plot is illustrated in Fig. 3.

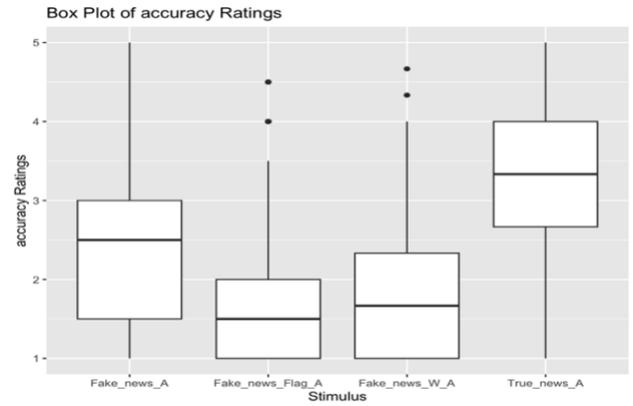

Fig. 3. Accuracy rating (where 1- Not accurate at all to 5 -Very accurate) for the different types of Stimuli.

*2) Sharing*

We also find that participants would share true news more compared to all the other three conditions. As seen with the accuracy ratings, we also find that fake news headlines attached with flags do reduce the sharing intent compared with no flags. However, there was no significant difference in sharing intent found between the flagging conditions. This indicates that even a subtle indication that a news item is false will decrease individual intent to share it within their own circle. However, regardless of whether the news is true or false, an individual's intention to share is very unlikely. To illustrate the effects of the flags on the sharing intent, a box plot is illustrated in Fig. 4.

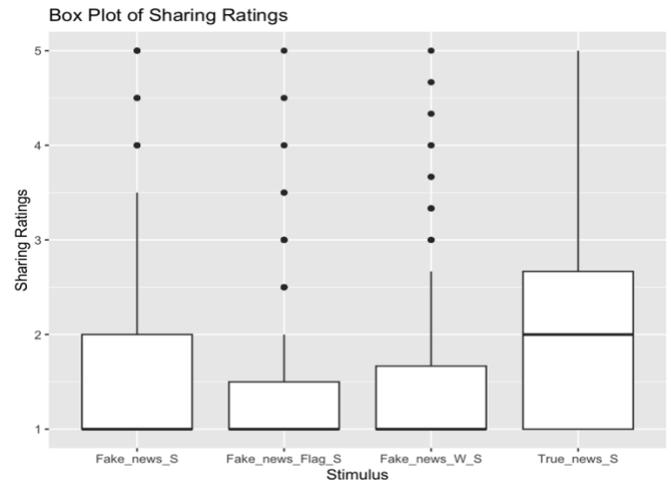

Fig. 4. Sharing intent (1 – Extremely Unlikely to 5 – Extremely likely) for the different types of Stimuli.

Interestingly, using the Mann-Whitney U test, we found that there is a significant difference between the sharing intent of males and females within our study. We found that males tend to share more compared to females, where males tend to share true news headlines, fake news headlines with warning flags and fake news headlines with warning and explanation flags within their social circle.

*3) Trust*

In the two flagging conditions, using the Friedman test, we find that there is a significant difference between the trustworthiness rating of the two flagged conditions (p-value = 0.0002602). Using the Wilcoxon signed-rank test, we find individuals report higher trustworthiness for the flag with the explanatory text compared to just the flag indicating that the extra context added does increase the overall self-reported trustworthiness of the flag.

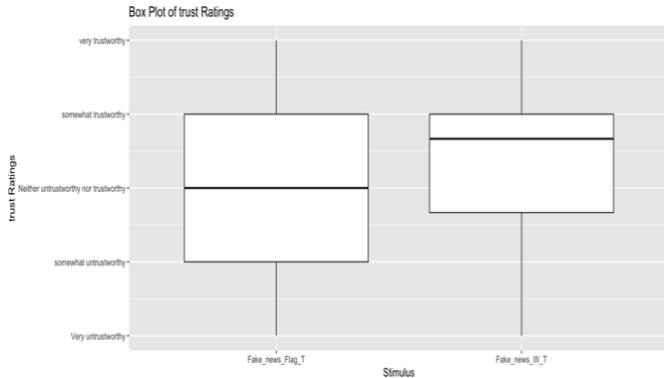

Fig. 5. Trustworthiness rating (where 1- Very untrustworthy to 5 -Very trustworthy) for the two flagging stimuli, where Fake_News_Flag_T denotes the trustworthiness of Fake news headlines with a warning flag and Fake_News_W_T denotes the trustworthiness of fake news headlines with a warning and explanation flag.

*4) Correlations with independent variables*

In our correlation analysis, we observed several significant correlations among a range of variables including age, education, social media use, political labelling, and various factors related to the perceived accuracy, sharing of true and fake news headlines and trust of flags.

For instance, we found a negative correlation between age and social media use (r = -0.18), indicating that the younger population is more likely to engage on social platforms. We also found a negative correlation between age and perceived fake news headline accuracy (r = -0.19) (fake headlines without any flags), suggesting that younger individuals are more susceptible to fake news headlines. This is reinforced by a negative correlation between age and fake news-sharing intent (r = -0.12), which indicates that younger individuals tend to share fake news headlines within their circles.

A particularly interesting finding emerged around the intersection of news sharing and social media use. We found a significant positive correlation between social media and fake news sharing (r = 0.18) and true news sharing (r = 0.21), implying that individuals who spend more hours on social media tend to share news headlines online. Additionally, we also found a positive correlation between social media use and the sharing intent of fake news with warning and explanation flags (r = 0.19), however, no significant correlation was found with the sharing intent of fake news headlines with just flags. This indicates that individuals who spend more time on social media tend to share fake headlines and would also share the context of why this headline is false within their social circle.

Interestingly, political ideology appears to significantly influence the perceived accuracy rating of fake news headlines across various forms. It significantly correlates positively with the perceived accuracy of fake news headlines (r = 0.21), flagged fake news headlines (r = 0.39), and flagged fake news headlines with explanatory texts (r = 0.25). This suggests that political ideology, particularly conservating leaning tends to judge fake news headlines as more accurate. We also find that there is a significant negative correlation between political ideology and perceived true news headline accuracy, suggesting that left-leaning individuals rate true news as more accurate compared to conservatives. In terms of sharing intent, political ideology exhibits a significant positive correlation with the sharing of fake news headlines (r = 0.16), flagged fake news headlines (r = 0.26) and flagged fake news headlines with explanatory texts (r = 0.13), and a negative correlation with sharing true news headline (r = -0.12). These correlations suggest that conservative-leaning individuals might be more likely to share both unmarked and flagged fake news while being less likely to share true news with their family and friends.

Among all the independent variables, we found that political ideology demonstrated a negative correlation with the trustworthiness of both flagging conditions (r = -0.34 for fake news with a warning flag and r = -0.37 for fake news with a warning and explanation flag). This indicates that conservative-leaning individuals generally have low trust in the fact-checkers or flagging system on social media. A correlation heatmap is given in Fig. 6.

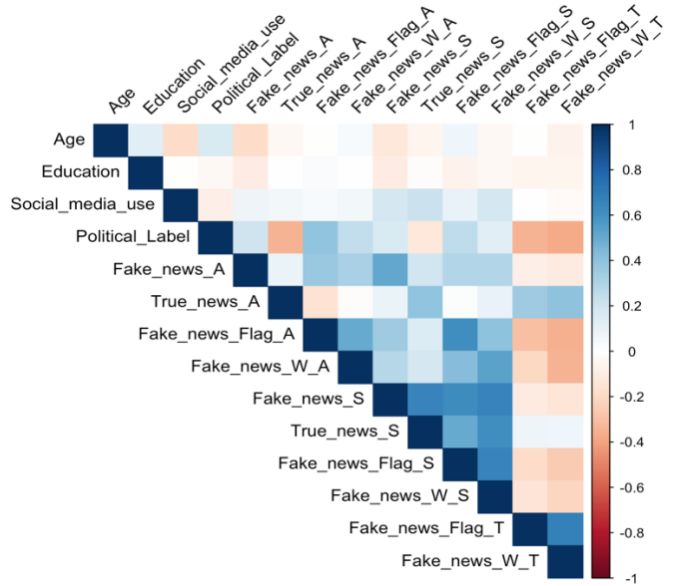

Fig. 6. A correlation heatmap between the independent and dependent variables. The four types are true news headlines (true_news), fake news headlines (fake_news), fake news headlines with warning flag(fake_news_Flag), and fake news headlines with warning and explanation flag (fake_news_W). Here, A denotes accuracy rating, S denotes sharing intent, and T denotes trustworthiness.

V. CONCLUSION AND FUTURE WORKS

In this paper, we discuss the effects of adding context to misinformation flags directly from fact-checking websites as a part of designing the flag. Our results echo the growing body of empirical work supporting the findings that fake news flags provide effective counter misinformation on social media [5], [6], [24]. More importantly, our results suggest that the absence of the warning text has a significant impact on participants' trustworthiness ratings of the flags. We find that participants rate flags with explanatory labels as more trustworthy than flags without them. Moreover, we found notable differences in accuracy ratings between fake news headlines with a flag and fake news headlines with an explanatory text, indicating a confounding effect on how the explanatory text should be incorporated into the design of misinformation flags. We also find that participants in general are extremely unlikely to somewhat unlikely share any headlines in the stimuli on social media or with friends or family. This is in line with the results from [25], where the authors found that generally individuals are reluctant to share information online.

Our study also found significant correlations among variables such as age, education, social media use, political ideology, and perceived news accuracy, news-sharing intent, and trust in warning flags. Younger individuals showed a higher susceptibility to fake news and a tendency to share it within their circles. This is in line with the findings of [22], [26], where the authors found that younger individuals tend to be susceptible to misinformation online.

We also echoed similar results from [27], that social media users are inclined to share news irrespective of their veracity, and those that spend more time on these platforms also tend to share flagged fake news with explanations. Political ideology greatly influenced perceptions of fake headline accuracy, with conservative-leaning individuals tending to view fake news as more accurate and sharing such news more often. Furthermore, these conservative-leaning individuals exhibited a lower trust in fact-checkers or flagging systems on social media. These insights highlight the intricate dynamics of news perception and sharing in the context of fake news, social media use, and political alignment.

Also, we found that overall, the participants in our study were very unlikely to share any sort of news within their own circle. This may be due to the increased awareness of the spread of misinformation on social media; many people have become more cautious and skeptical about online content. They might hesitate to share content, especially related to sensitive topics like health or politics, to avoid unintentionally spreading misinformation.

While our study uncovers valuable insights, there are a number of limitations. Firstly, we adopted an imitation of a social media experience, which reduced the response options usually provided by digital platforms, such as like and dislike buttons. Secondly, our survey was self-reported, potentially introducing social desirability bias, wherein participants may respond in a way they perceive as socially acceptable rather than reflecting their true behaviour. Furthermore, self-reported data often rely on the participant's subjective interpretation of questions, which could lead to variations in understanding and subsequently, inconsistency in responses. Thirdly, another limitation is the focus on COVID-19 related headlines in our experiment. While these headlines are timely and relevant, they may also elicit strong emotional reactions, possibly skewing participants' responses. There might be different implications on how our results might translate to less emotionally charged topics [28].

Despite the potential limitations, our study sheds valuable light on the effectiveness of incorporating context into misinformation flags on social media platforms. We offer evidence that such context can enhance the trustworthiness of these flags and improve users' judgment of news accuracy. These findings have significant implications for the design and implementation of counter-misinformation strategies. Future research should continue to build on these findings, investigating other relevant factors and refining the design of misinformation flags for optimal impact. One such direction would be to investigate personalisation in this field of research. Research in the field of persuasive technologies [29], [30] has indicated that personalised approaches to individuals provide a better response for persuasion than a "one size fits all" type solution. As stated in [31], personal efficacy is one of the reasons why an individual reacts to fake news. Therefore, it would be worthwhile to investigate whether different designs (whether a different format for explanation text or the design) for misinformation flags would increase trustworthiness among user groups such as different age groups, genders, and other moderating factors. It would also be worthwhile investigating explanations by AI systems and adding them as context to how and why these AI systems flagged content as misinformation.

ACKNOWLEDGMENT

This work was conducted with the financial support of the Science Foundation Ireland Centre for Research Training in Digitally-Enhanced Reality (D-real) under Grant No. 18/CRT/6224, the VIGILANT project that has received funding from the European Union's Horizon Europe Programme under Grant Agreement No. 101073921 and at the ADAPT SFI Research Centre at Trinity College Dublin. ADAPT, the SFI Research Centre for AI-Driven Digital Content Technology is funded by Science Foundation Ireland through the SFI Research Centers Programme and is co-funded under the European Regional Development Fund (ERDF) through Grant Agreement No. 13/RC/2106.